# Fully Automated OCT-based Tissue Screening System

Shaohua Pi,[1,2,*] Razieh Ganjee,[1] Lingyun Wang,[1,2] Riley K. Arbuckle,[1,3,4] Chengcheng Zhao,[1,2] Jose A Sahel,[1] Bingjie Wang,[1] and Yuanyuan Chen[1,3]

[1]Department of Ophthalmology, University of Pittsburgh, Pittsburgh, PA 15213, USA
[2]Department of Bioengineering, University of Pittsburgh, Pittsburgh, PA 15213, USA
[3]Department of Pharmacology and Chemical Biology, University of Pittsburgh, Pittsburgh, PA 15213, USA
[4]Department of Human Genetics, University of Pittsburgh Graduate School of Public Health, Pittsburgh, PA 15213, USA

*shaohua@pitt.edu

***This study introduces a groundbreaking optical coherence tomography (OCT) imaging system dedicated for high-throughput screening applications using ex vivo tissue culture. Leveraging OCT's non-invasive, high-resolution capabilities, the system is equipped with a custom-designed motorized platform and tissue detection ability for automated, successive imaging across samples. Transformer-based deep learning segmentation algorithms further ensure robust, consistent, and efficient readouts meeting the standards for screening assays. Validated using retinal explant cultures from a mouse model of retinal degeneration, the system provides robust, rapid, reliable, unbiased, and comprehensive readouts of tissue response to treatments. This fully automated OCT-based system marks a significant advancement in tissue screening, promising to transform drug discovery, as well as other relevant research fields.***

*http://doi.org/10.1364/OL.XXXXXX*

Screening systems, especially in drug discovery, are crucial for identifying potential therapeutic compounds from a vast array of chemicals, significantly expediting the progression of hit identification[1]. By offering a unique balance between the simplicity of cell-based assays and the complexity of *in vivo* animal models, *ex vivo* tissue culture is emerging for screening to provide a more physiologically relevant context in studying drug effects and disease mechanisms. However, the lack of an effective *ex vivo* imaging tool limits its usage to assess the biological activity, efficacy, and safety of substances in a high-throughput manner.

Optical coherence tomography (OCT)[2] has long been a cornerstone in ophthalmological diagnostics due to its non-invasive, high-resolution, and 3-D imaging capabilities. Unlike traditional histology—which is time-consuming, structurally disruptive, and often necessitates multiple samples—OCT may offer a rapid, longitudinal, and unbiased approach to quantitatively characterize tissue morphological responses to variables such as pharmacological agents. However, despite its clinical prominence, the role of OCT in *ex vivo* applications has been understudied, partially due to the lack of dedicated customization and optimization.

In this study, we report a fully automated OCT-based tissue screening system for research using *ex vivo* tissue culture, especially for applications requiring high-throughput screening of therapeutic agents, potential therapies, and optimal protocols. The unique features of the system include automated and successive OCT scan acquisition, as well as automated and reliable parameter readout, which are empowered with both hardware integrations (motorized platform with object detection ability) and algorithm (vision transformer for segmentation) innovations.

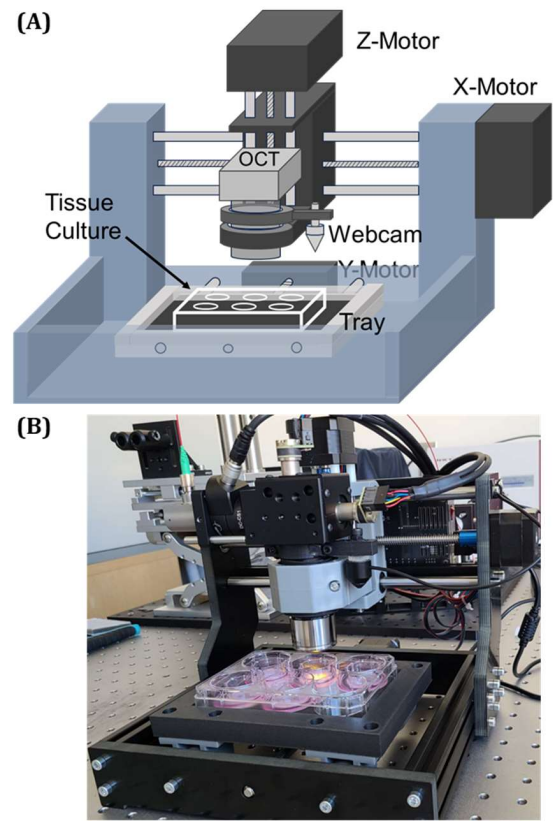

Fig. 1. (a) Schematic diagram and (b) a photograph of custom-designed motorized platform for the automated OCT scan acquisition.

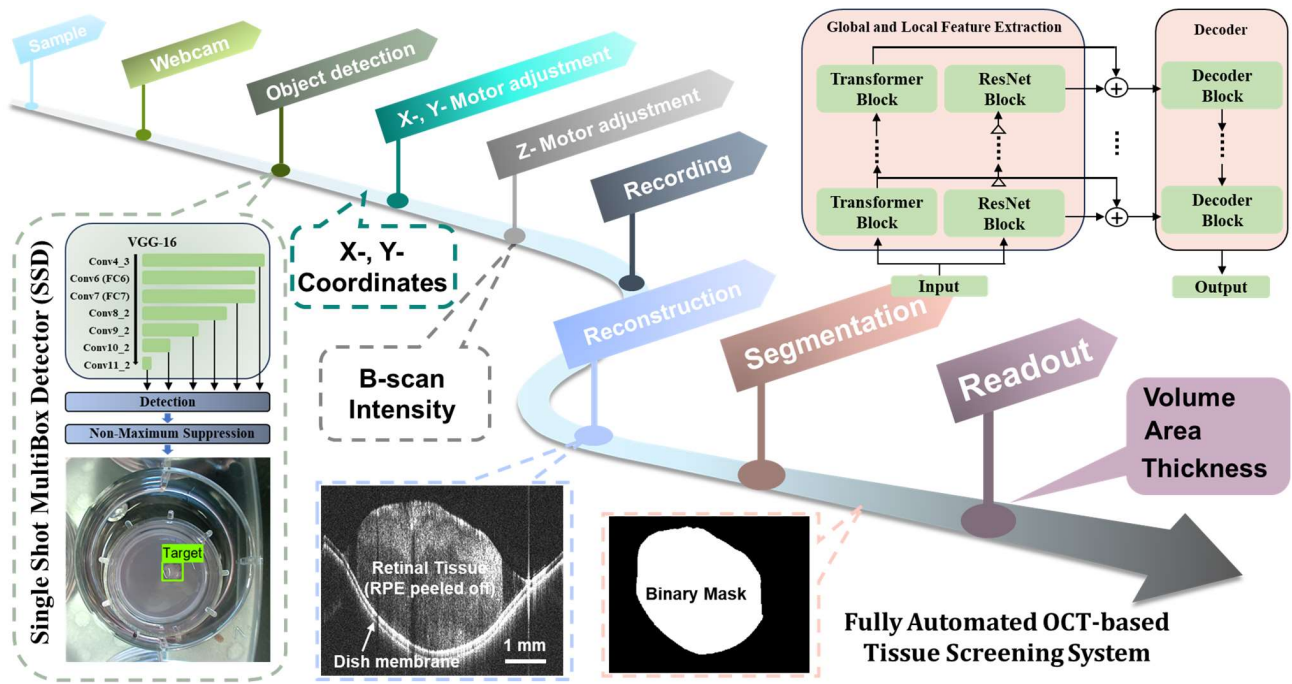

Fig. 2. Pipeline of the fully automated OCT-based tissue screening system.

The first bottleneck of tissue screening systems is the evaluation of numerous tissue samples quickly and reliably, as conventional histological imaging is prohibitively complicated and time-consuming for screening purposes. To address this issue, we have designed and customized a motorized platform enabling 3-D manipulation for automated and successive OCT imaging across tissue samples (Fig. 1). This platform comprises three motorized linear stages installed in the X-, Y-, and Z- directions, respectively. The tissue culture plate is situated in a tray with dimensions carefully tailored to fit a standard multi-well culture plate (86×128 mm) and is installed in the Y- linear stage slider. The OCT sample arm is mounted on the Z- linear stage slider, while the entire top structure is mounted to X- linear stage slider, allowing the OCT probing light beam to be manipulated and moved within the X-Z plane. Additionally, a webcam is mounted parallel to the OCT sample arm, with its field of view (FOV) adjusted to cover an area slightly larger than a single well. Once installed, the webcam's position is precisely measured and calibrated with the OCT probe (X- and Y- distances) to guide OCT imaging.

The pipeline of the screening system is summarized in Fig. 2. Before imaging, the position of reference arm, beam focus and polarization of OCT is pre-optimized. At initial status, the X- and Y- linear stages position the webcam at the center of the first well (A1: first row and first column) for a tissue culture plate. A picture of the well is then captured by the webcam and processed by an object detection algorithm to identify the presence and precise location of any tissue sample. The output of the object detection is a bounding box containing the target tissue. The centroids of the sample, i.e., the X- and Y- coordinates are converted to corresponding voltages to adjust the X-, and Y- motors, thereby and driving the OCT light beams to the tissue area.

Once the OCT light beam is horizontally aligned with the tissue area, the position of Z-motor is swept under M-B scan mode to find the optimal optical length for the OCT sample arm. The optimal depth is determined by examining the intensity averaged across the entire B-scan. The Z- motor position with the greatest brightness is noted and used later to trigger the saving module for volumetric OCT scan recording. From our experimental experience, the optimal Z- motor position remained consistent among samples. Therefore, the sweeping procedure can be skipped within one culture plate or limited to a narrow range during practice. After completing the scan acquisition, the B-scan images are segmented by a pre-trained deep-learning network to delineate the tissue regions. Thickness (μm), area ($mm^2$) and volume ($mm^3$) are calculated as readouts from the segmented binary images.

The performance of this system was validated for drug screening using retinal explant cultures. The retinal explants were prepared from the retinae from $Rho^{P23H/+}$ mice, isolated at P15 (day 0 in culture)[3]. The $Rho^{P23H/+}$ mouse model of retinitis pigmentosa is characterized by progressive thinning[4] of the photoreceptor layers due to rod cell death. The total thickness of the retina explant from this mouse model decreases significantly within 10 days of culture (P25) compared to the wildtype control[5], providing sufficient morphological changes detectable by OCT and amenable to rescued by efficacious drugs. Six-well plates were used in this study, with each well containing one whole retina explant. Two groups of samples were created by administering negative control (or vehicle control) and positive control compounds. OCT scans (FOV:6×6 mm, A-lines: 500×2×500) were acquired using a visible light OCT prototype[6] at baseline and day 10 in culture to examine the efficacy of treatment.

The implementation was carried out using the Pytorch framework on a PC with an i9-12900K CPU, NVidia GeForce RTX 3080 Ti, and 64 GB of RAM. For object detection, we developed a deep learning algorithm based on the innovative Single Shot MultiBox Detector (SSD)[7]. The advantage of SSD is that it maintains accuracy while significantly improving processing speed by eliminating proposal generation and subsequent pixel or feature resampling stages, encapsulating all computation in a single network. Convolutional feature layers with progressively decreased size were added to the truncated base network, allowing prediction of detection at multiple scales (Fig. 2). Three hundred webcam images of the wells with a FOV of 30×30-mm were acquired and manually labeled for training (240 images) and testing (60 images). A weighted sum of the localization loss and the confidence loss was utilized during training[7]. We used a batch size of 4 and stochastic gradient descent (SGD) optimizer with an initial learning rate of 5e-4. The centroid of tissue sample was determined within 7 milliseconds. The mean average precision (with an Intersection Over Union threshold set at 0.5 to 0.95 with an interval of 0.05) was 0.889, indicating precise location detection. The centroid shift between the predictions and the labels was ~10 µm. The success rate of detection was 100%, with void output for all empty wells, indicating the high reliability of this algorithm.

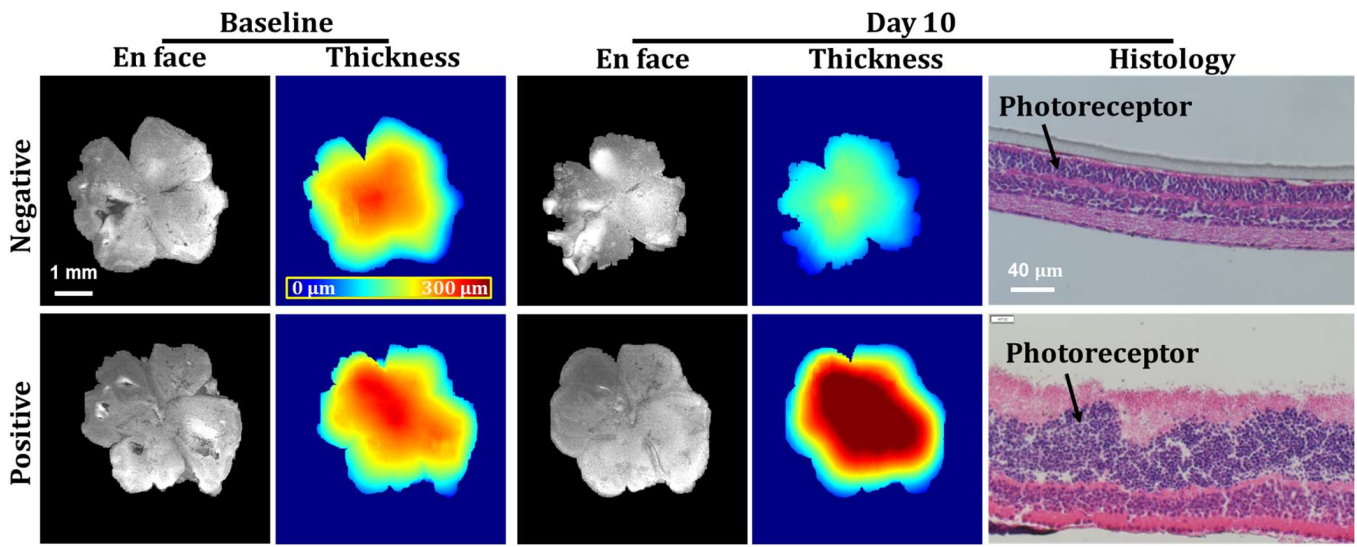

Fig. 3. The OCT *en face* image and thickness map for $Rho^{P23H/+}$ mice retina at Baseline (P15) and Day 10 (P25) treated with negative and positive compounds. Histology images were obtained after Day 10 screening for validation.

The challenges during segmentation include frequent artifacts of specular reflection, inconsistent tissue reflectance, and strong interference from the tissue-supporting membrane. To overcome these challenges, we developed a transformer-based method employing a hybrid architecture of ResNet[8] and multi-scale hierarchical transformer[9] to efficiently capture both the local and global features. As shown in Fig. 2, the backbone of the model comprised two independent networks: multipath transformer networks for extracting long-distance dependencies of explant features and a residual network for extracting local features. OCT B-scan images were first processed by these two backbones in multiple scales in parallel. The local and global features provided by these two networks were then concatenated and sent to the corresponding decoder block to produce the segmentation map. For tissue segmentation, we used 2150 manually labeled B-scan images for training. Data augmentation, including translation, contrast adjustment, and vertical/horizontal flip was applied to enlarge the training set. We utilized the learning rate of 1e-2, a batch size of 32, epochs number of 30, and the AdamW optimizer. The recall, precision, and dice rates of the segmentation model were 0.87±0.20, 0.84±0.25, and 0.89±0.24, respectively. The processing time for each B-scan image was around 150 milliseconds. These parameters suggest that the proposed algorithm provides a fast and accurate segmentation performance.

The resulting *en face* images, including the structure and thickness heatmap were shown in Fig. 3. Tissue area and volume were calculated by projecting and integrating all B-scans. To avoid bias from the periphery, the average thickness was calculated only in the central tissue regions, where the tissue was flatter with thickness larger than the median value. For characterization, we programmed the system to take five volumes from each sample. Repeatability was calculated as the pooled standard deviation among the repetitions for all samples. The repeatability was high for all readouts (volume: 0.005 $mm^3$, area: 0.039 $mm^2$, thickness: 0.4 µm), indicating highly reliable measurements offered by this system. The reproducibility, calculated as the pooled standard deviation among the samples under the same conditions, was 0.107 $mm^3$ for volume, 0.755 $mm^2$ for area, and 12.1 µm for thickness, all corresponding to less than ~5% variation of the measurements.

We found that the $Rho^{P23H/+}$ mouse retina thickness was significantly (~30% or 60 µm) reduced at day 10 in culture (133.7 µm ± 8.4 µm) compared to baseline (191.8 µm ± 11.5 µm) in the negative control treatment group due to progressive retinal degeneration (Fig. 4), consistent with our previous report[5]. Consequently, the area of retinal explant, uniquely captured by OCT due to its wide-field volumetric imaging, also demonstrated a slight shrinkage (~20%) during the monitoring period, possibly due to the shear force induced by the tissue thinning. Additionally, tissue volume reduction of approximately 1 $mm^3$ also characterized the retinal degeneration. In contrast, the positive treatment not only ceased the thinning but also led to an approximately 30% increase in thickness and volume (Fig. 4). Meanwhile, the area didn't show changes in the positive control group. The protective effect of the positive treatment was further validated by histology with a thicker retina and preserved photoreceptor cell number (Fig. 3).

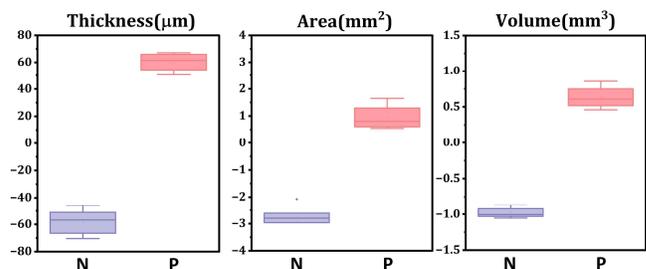

Fig. 4. Changes of tissue thickness (left), area (middle) and volume (right) from baseline to day 10 in culture for both negative (N=4) and positive (N=8) treated groups. The middle lines and treat error bars are means ± SDs. P-values<0.05 for all readouts between the groups.

Next, we computed two metrics[10], the Z-prime factor (Z') and the strictly standardized mean difference (SSMD, denoted as β) which are commonly used to assess assay quality for drug screening (Eq. 1). The μ and σ were the mean and standard deviation values from the negative and positive control groups respectively. The Z'-factor describes the distinction between positive and negative controls with a value between 0.5 and 1 being considered excellent and recommended for a plate-reader-based drug screening assay. In our study, the Z'-factors were calculated as 0.70 for volume measurement and 0.55 for thickness measurement, both of which are exceptional scores for such a complex tissue-based imaging assay. The SSMD measures the strength of the difference between the controls. Acceptable values for SSMD depend on the strength (effect size) of the positive control. For the sample size of controls in our study, a β value larger than 2 is considered excellent. The β values were calculated as 8.8 and 9.3 for volume and thickness respectively.

$$Z' = 1 - \frac{3\sigma_{positive} + 3\sigma_{negative}}{|\mu_{positive} - \mu_{negative}|}$$

$$\beta = \frac{\mu_{positive} - \mu_{negative}}{\sqrt{\sigma_{positive}^2 + \sigma_{negative}^2}}$$

(1)

These evaluations demonstrate that our system holds translational potential for tissue-based drug screening. Retinal tissue culture is more advantageous than cell culture as it provides a drug testing platform containing a multi-cellular network. For example, the pathophysiology of retinal degeneration is complex and involves crosstalk between the photoreceptors, the inner retinal neurons, the Müller glia and microglia[11]. When dissociated and cultured alone, photoreceptors quickly become round cells and lose most of their biological features observed in the tissue context. The retinal explant culture can maintain the retinal layers for up to four weeks, providing a physiologically relevant retinal degeneration model similar to that observed *in vivo*, yet allowing drug screening in a multi-well format without concerns of the drug bioavailability *in vivo*.

This automated imaging system for *ex vivo* tissue can lead to a more accurate evaluation of efficacy and toxicity of hits, therefore shortening the drug discovery pipeline and increasing the success rate of finding efficacious drugs for retinal degeneration. Base-line measurements and whole tissue volumetric quantification normalize sample-to-sample variations. More importantly, tissue-based assays are faster, less expensive, and circumvent ethical concerns associated with animal research. They allow for the examination of drug responses and safety directly on human postmortem tissue explants, which are often not replicable in animal models. This approach bridges the gap between animal studies and clinical applications, enhancing the translational relevance of preclinical research.

Although validated using a 6-well plate, our setup can be easily adapted for a 48- or 96-well plate (subject to the optimization of explant culturing protocol) with each well culturing one/multiple portions of retina biopsies. This means the retina tissue from one animal/donor could provide tissue biopsies for testing multiple conditions in the future. The testing agents are not limited to small molecule drugs but can include peptides, antibodies, adeno-associated viral vectors, small RNAs, etc. Other applications, such as organoid screening and gene therapy development for vision research, as well as a wide range of research including developmental biology studies, cancer research, and toxicity testing, might benefit from the promise of this system. By providing 3-D morphological tissue responses to various treatments, this platform may accelerate the therapeutic development pipeline by screening for tissue-protecting agents.

The current limitation of manually isolating the tissues and placing plates into the system represents a bottleneck in scaling up the technology for broader applications. Plans for an updated version include automating plate loading and unloading, significantly enhancing throughput and reducing the potential for human error, thereby streamlining the workflow for large-scale screenings. Future directions could also include integrating artificial intelligence for real-time data analysis, predictive modeling of treatment outcomes, and developing a more compact, user-friendly design to facilitate adaption in standard laboratory settings. Such innovations would further unlock the potential of this technology in basic science and therapeutic development. Technical improvements of tissue biopsy isolation are also underway to improve the efficiency of the screening system.

In summary, we developed an innovative OCT-based tissue screening system tailored for high-throughput applications in drug discovery and evaluation, providing multiplex readouts for unbiased tissue morphological description for the first time. Key innovations include a custom-designed motorized platform for precise 3-D positioning of samples, alongside sophisticated deep learning algorithms for automated tissue detection and segmentation. Validated through rigorous testing with retinal explant cultures, our system demonstrated its efficiency in accurately measuring drug efficacy, overcoming the limitations in the traditional histology methods. This study not only showcases the system's potential to streamline preclinical vision research but also highlights its applicability across a spectrum of biomedical fields.

**Acknowledgements.** Eye and Ear Foundation, NIH (R01 EY030991, P30 EY08098), Research to Prevent Blindness.

**Disclosure.** The authors declare no conflicts of interest.

**References**
1. Bray, M.-A. & Carpenter, A. (2017) *Assay Guidance Manual [Internet]*.
2. Huang, D. et al. (1991) *Science* **254**, 1178-1181.
3. Sakami, S. et al. (2011) *Journal of Biological Chemistry* **286**, 10551-10567.
4. Wang, B. et al. (2024) *Experimental Eye Research*, 109826.
5. Vats, A. et al. (2022) *JCI insight* **7**.
6. Wang, B. et al. (2024) *Biomedical Optics Express* **15**, 3112-3127.
7. Liu, W. et al. in Secondary, Vol. Computer Vision–ECCV 2016 21-37 (Springer, 2016).
8. He, K., Zhang, X., Ren, S. & Sun, J. in Secondary 770-778 (2016).
9. Lee, Y., Kim, J., Willette, J. & Hwang, S.J. in Secondary 7287-7296 (2022).
10. Zhang, J.-H., Chung, T.D. & Oldenburg, K.R. (1999) *Journal of biomolecular screening* **4**, 67-73.
11. Massengill, M.T., Ash, N.F., Young, B.M., Ildefonso, C.J. & Lewin, A.S. (2020) *Sci Rep* **10**, 16967.